\documentclass[
reprint,
amsmath,amssymb,
aps,
pra,
superscriptaddress,
]{revtex4-2}

\usepackage{afterpage}
\usepackage{adjustbox}
\usepackage{rotating}
\usepackage[sort&compress]{natbib}
\usepackage{graphicx}
\usepackage{dcolumn}
\usepackage{bm}
\usepackage[colorlinks=true, urlcolor=blue, pdfborder={0 0 0}]{hyperref}
\hypersetup{
linkcolor=blue,
citecolor=blue
}

\begin{document}

\preprint{APS/123-QED}

\title{Full Band Structure Calculation of Semiconducting Materials on a Noisy Quantum Processor}

\author{Shaobo Zhang}
\email{shaozhang@student.unimelb.edu.au}
\affiliation{School of Physics, The University of Melbourne, Parkville, 3010, Australia}

\author{Akib Karim}
\affiliation{Data61, CSIRO, 3168, Clayton Australia}

\author{Harry M. Quiney}
\affiliation{School of Physics, The University of Melbourne, Parkville, 3010, Australia}

\author{Muhammad Usman}
\affiliation{School of Physics, The University of Melbourne, Parkville, 3010, Australia}
\affiliation{Data61, CSIRO, 3168, Clayton Australia}

\date{\today}

\begin{abstract}

Quantum chemistry is a promising application in the era of quantum computing since the unique effects of quantum mechanics that take exponential growing resources to simulate classically are controllable on quantum computers. Fermionic degrees of freedom can be encoded efficiently onto qubits and allow for algorithms such as the Quantum Equation-of-Motion method to find the entire energy spectrum of a quantum system. In this paper, we propose the Reduced Quantum Equation-of-Motion method by reducing the dimensionality of its generalized eigenvalue equation, which results in half the measurements required compared to the Quantum Equation-of-Motion method, leading to speed up the algorithm and less noise accumulation on real devices. In particular, we analyse the performance of our method on two noise models and calculate the excitation energies of a bulk Silicon and Gallium Arsenide using our method on an IBM quantum processor. Our method is fully robust to the uniform depolarizing error and we demonstrate that the selection of suitable atomic orbital complexity could increase the robustness of our algorithm under real noise. We also find that taking the average of multiple experiments tends towards the correct energies due to the fluctuations around the exact values. Such noise resilience of our approach could be used on current quantum devices to solve quantum chemistry problems.

\end{abstract}

\maketitle


\section{\label{sec-Introduction}Introduction} 

Solving the many-body electronic Hamiltonian accurately for molecular systems and materials from first principles is crucial in condensed matter physics, which has been approximated and addressed by numerical algorithms over the last few decades\cite{szabo2012modern,helgaker2013molecular,bartlett2007coupled,mcardle2020quantum}. However, such problems can quickly become intractable for classical computers due to their limited electronic structure descriptions as the system size grows. Contrarily, quantum computers could efficiently solve many-electronic structure problems using the properties of quantum mechanics\cite{feynman2018simulating,nielsen2010quantum}, but the presence of noise or errors in quantum processors greatly affects the calculation results. Therefore, the development and implementation of noise-robust quantum algorithms are important to enhance the efficacy of the current generation of quantum computers in the era of Noisy Intermediate-Scale Quantum\cite{preskill2018quantum} (NISQ) devices.

The Variational Quantum Eigensolver\cite{peruzzo2014variational, o2016scalable, kandala2017hardware, google2020hartree} (VQE) has shown success for ground state calculations of small molecular systems, which has inbuilt robustness to coherent errors and can achieve high accuracy calculations with additional quantum error mitigating methods\cite{nigg2014quantum, mcclean2017hybrid, li2017efficient, bonet2018low, temme2017error, endo2018practical, kandala2017hardware, kandala2019error, takagi2021optimal, nation2021scalable, urbanek2021mitigating, urbanek2020error, guo2022quantum, yoshioka2022generalized, karim2024low}. Beyond the ground state calculation, obtaining the full band structure of materials requires excited states. Recently, a number of hybrid quantum-classical approaches have been designed as extensions of VQE to obtain the excited state properties of the small molecular systems on simulators\cite{higgott2019variationalquantum, mcclean2017hybrid, asthana2023quantum, xie2022orthogonal, russo2021evaluating, bauman2020toward, sugisaki2021bayesian} and noisy quantum devices\cite{colless2018computation, ollitrault2020quantum, ganzhorn2019gate, sureshbabu2021implementation}. For semiconducting materials, the work published in the literature has demonstrated the calculation of the incomplete band structure of Silicon (Si) on noisy quantum hardware\cite{cerasoli2020quantum,ohgoe2024demonstrating}, and the full band structure of periodic systems on simulators\cite{nakanishi2019subspace, fan2021equation, dittmer2019accurate, sureshbabu2021implementation, clinton2024towards} with only additional measurements on the VQE circuit, but to date these methods have not calculated the full band structure on a real device. Such methods are susceptible to incoherent noise and extra measurements will increase the effect of noise.

\begin{figure*}
\includegraphics[scale=0.53]{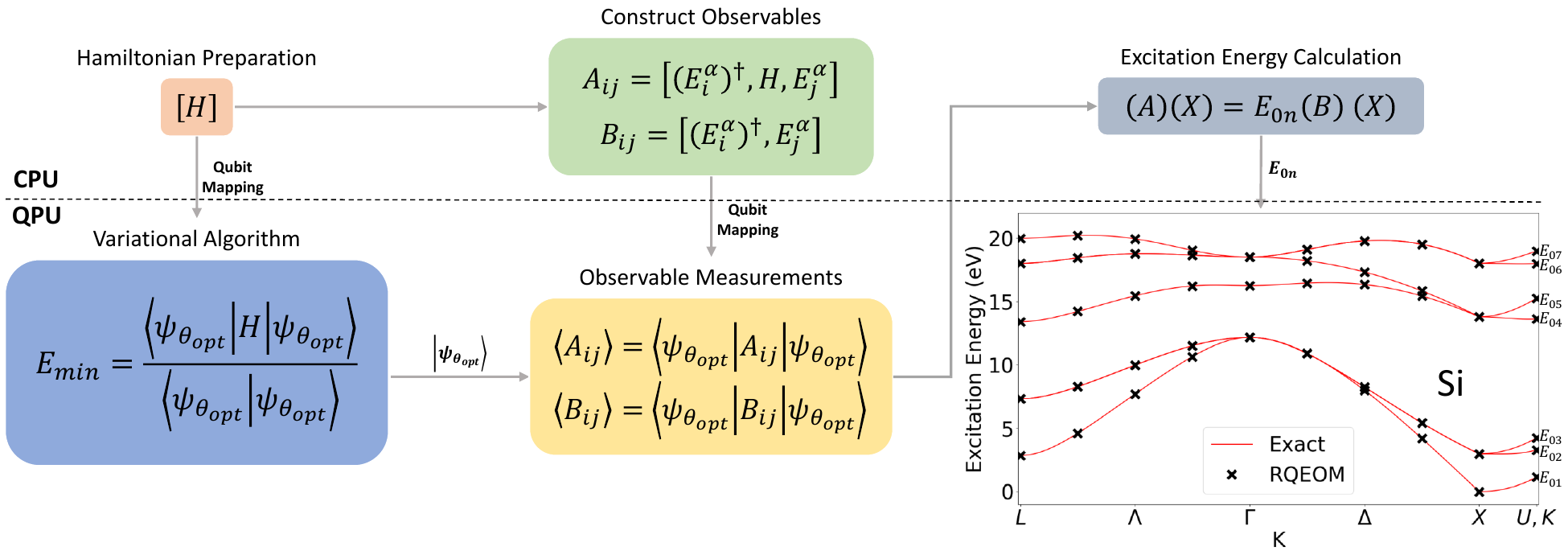}
\caption[The whole procedure of RQEOM method]{
\label{fig1} 
A flowchart diagram that illustrates the procedure of RQEOM approach. The colored areas describe each step of the algorithm, the figure after the blue-grey rounded rectangle represents the numerical simulation of the excitation energy of Si using RQEOM method, and the red line denotes the exact excitation energy calculated by the classical diagonalization of the Hamiltonian and the black crosses depict the excitation energies calculated by RQEOM method. The rounded rectangles above the dashed line are calculated on the classical processor unit (CPU) and the rounded rectangles under the dashed line are measured on the quantum processor unit (QPU).
}
\end{figure*}

In this paper, we propose a new method henceforth labeled as the Reduced Quantum Equation-of-Motion (RQEOM), a variance of the Quantum Equation-of-Motion\cite{ollitrault2020quantum} (QEOM) method which halves the required observables by neglecting the de-excitation operators. The procedures of the RQEOM method are summarized as a flow chart diagram in Fig.~\ref{fig1}. we illustrate the theoretical foundation of this work in Section~\ref{sec-Theory}. In Section~\ref{subsec-Experiment-Simulation}, We investigate the effect of incoherent noise on our method and find it is immune to the uniform depolarising noise, as such, we further investigate its performance under a more realistic noise model, the biased depolarising noise\cite{aliferis2009fault, nigg2014quantum, robertson2017tailored, tuckett2020fault, tuckett2018ultrahigh, gicev2023quantum}, and find that the RQEOM method demonstrates strong resilience. We utilize this inherent noise mitigation to compute the full band structure of two semiconducting materials Si and Gallium Arsenide (GaAs) on a noisy superconducting quantum chip without any noise mitigation.


\section{\label{sec-Theory}Theory}

\subsection{\label{subsec-EOM}Quantum Equation-of-Motion Method}

Given a Hamiltonian $H$ of a quantum system, the wavefunction $|\psi_n\rangle$ and the corresponding energy $E_n$ can be obtained by solving the time-independent Schr\"odinger equation 
\begin{eqnarray}
\label{eq-tischrodingereq}
H | \psi_n \rangle = E_n | \psi_n \rangle ,
\end{eqnarray}
where the subscript $n$ indicates $nth$ excite state, $n=0$ is the ground state.

The EOM method was firstly derived by Rowe\cite{rowe1968equations} to calculate the energy gaps between the ground state energy and each excited state energy for classical simulations, which was then generalized as one of the approaches to obtain the excited state energies of quantum systems for quantum computers \cite{ollitrault2020quantum,fan2021equation,ganzhorn2019gate}. The EOM method first constructs a super excitation operator $O_n^\dagger$ which excites the ground state to $nth$ excited state and a super de-excitation operator $O_n$ which de-excites the $nth$ excited state to the ground state respectively,
\begin{eqnarray}
\begin{aligned}
\label{eq-superoperatorO}
O^{\dagger}_n | \psi_0 \rangle &= | \psi_n \rangle, \\
O_n | \psi_n \rangle &= | \psi_0 \rangle.
\end{aligned}
\end{eqnarray}
The energy gap $E_{0n} = E_n - E_0$. Substituting Eq.~(\ref{eq-tischrodingereq}) and Eq.~(\ref{eq-superoperatorO}) into $E_{0n}|\psi_n\rangle$, we can obtain
\begin{eqnarray}
\begin{aligned}
E_{0n}|\psi_n\rangle
&= (E_n - E_0)O^{\dagger}_n|\psi_0\rangle \\
&= HO^{\dagger}_n|\psi_0\rangle - O^{\dagger}_nH|\psi_0\rangle \\
&= [H, O_n^{\dagger}]|\psi_0\rangle,    
\end{aligned}
\end{eqnarray}
Multiplying $\langle\psi_0|O_n$ on both sides of the equation above, the energy gap $E_{0n}$ can then be expressed as
\begin{eqnarray}
\label{eq-qeom1}
E_{0n} = 
\frac
{\langle\psi_0|[O_n,[H, O_n^{\dagger}]]|\psi_0\rangle}
{\langle\psi_0|[O_n,O_n^{\dagger}]|\psi_0\rangle}.
\end{eqnarray}

Given the exact ground state 
$|\psi_0\rangle$, the commutator on the denominator has the relation $\langle\psi_0|[O_n,[H, O_n^{\dagger}]]|\psi_0\rangle = 
\langle\psi_0|[[O_n,H], O_n^{\dagger}]|\psi_0\rangle$ and the double commutator 
$
[O_n, H, O_n^{\dagger}] = 
\{[O_n,[H, O_n^{\dagger}]] + 
[[O_n,H], O_n^{\dagger}]\}/2.
$
Thus, Eq.~(\ref{eq-qeom1}) can be further written as 
\begin{eqnarray}
\label{eq-qeom2}
E_{0n} = 
\frac
{\langle\psi_0|[O_n, H, O_n^{\dagger}]|\psi_0\rangle}
{\langle\psi_0|[O_n,O_n^{\dagger}]|\psi_0\rangle}.
\end{eqnarray}

The superoperator $O^{\dagger}_n$ can be expanded as
\begin{eqnarray}
\label{eq-bigo}
O_n^{\dagger} = \sum_\alpha\sum_{i=0}^n(X_iE_i^\alpha - Y_i(E^{\alpha}_i)^{\dagger}),
\end{eqnarray}
where $\alpha$ indicates the type of excitation operator $E$, e.g., $\alpha = 2$ indicates the double excitation operator, $n$ is the number of excitation operators vary by $\alpha$, $X$ and $Y$ are the parameter vectors that span the coefficient space, $O_n$ is the complex conjugate of $O_n^\dagger$. By inserting the expanded superoperators $O^{\dagger}_n$ and $O_n$ into Eq.~(\ref{eq-qeom2}) and applying the linear variational principle by requiring $\delta E_{0n} = 0$, we can then solve the partial differential equation to obtain a generalized eigenvalue equation in the form of 
\begin{eqnarray}
\label{eq-qeom3}
\begin{pmatrix}
    \mbox{$M$}&\mbox{$Q$} \\
    \mbox{$Q^\ast$}&\mbox{$M^\ast$}
\end{pmatrix}
\begin{pmatrix}
    \mbox{$X$} \\
    \mbox{$Y$}
\end{pmatrix}
= 
E_{0n}
\begin{pmatrix}
    \mbox{$V$}&\mbox{$W$} \\
    \mbox{$-W^\ast$}&\mbox{$-V^\ast$}
\end{pmatrix}
\begin{pmatrix}
    \mbox{$X$} \\
    \mbox{$Y$}
\end{pmatrix},
\end{eqnarray}
where
\begin{eqnarray}
\begin{aligned}
\label{eq-mqvw}
\langle M_{ij} \rangle &=
\langle\psi_0|[(E^{\alpha}_i)^{\dagger},H,E^{\alpha}_j]|{\psi_0}\rangle, \\
\langle Q_{ij} \rangle &= 
-\langle\psi_0|[(E^{\alpha}_i)^{\dagger},H,(E^{\alpha}_j)^{\dagger}]|{\psi_0}\rangle, \\
\langle V_{ij} \rangle &= 
\langle\psi_0|[(E^{\alpha}_i)^{\dagger},E^{\alpha}_j]|{\psi_0}\rangle, \\
\langle W_{ij} \rangle &= 
-\langle\psi_0|[(E^{\alpha}_i)^{\dagger},(E^{\alpha}_j)^{\dagger}]|{\psi_0}\rangle.    
\end{aligned}
\end{eqnarray}
The entries of each matrix in above are the expectation values of the given observables $[(E^{\alpha}_i)^{\dagger},H,E^{\alpha}_j]$, $-[(E^{\alpha}_i)^{\dagger},H,(E^{\alpha}_j)^{\dagger}]$, $[(E^{\alpha}_i)^{\dagger},E^{\alpha}_j]$ and $-[(E^{\alpha}_i)^{\dagger},(E^{\alpha}_j)^{\dagger}]$ respectively, which will then be measured as expectation values of the ground state on quantum computers, a quantum system with $n$ pre-defined excitation operators gives $n$ excited state energy gaps and the $M$, $Q$, $V$ and $W$ matrices have $16n^2$ observables in total, each observable is a linear combination of Pauli terms. Eq.~(\ref{eq-qeom3}) will be solved classically and generate $2n$ solutions, $n$ of them have non-negative values and will be treated as energy gaps of the given $H$, since $E_{0n} \leq E_{0(n+1)}$, ranking $n$ positive values from the smallest to the largest would give the correct order of energy gaps; the remaining $n$ solutions are non-positive and will be regarded as the de-excitation energies\cite{ollitrault2020quantum} due to the symmetrization of Eq.~(\ref{eq-qeom3}).

\subsection{\label{subsec-RQEOM}Reduced Quantum Equation-of-Motion Method}

In QEOM method, obtaining the excitation energy gaps will automatically give the knowledge of the corresponding de-excitation energies, thus, Eq.~(\ref{eq-qeom3}) contains extra dimensions in vector space and requires more measurements on quantum computers. The eigenspace of a generalized eigenvalue equation grows exponentially as the quantum system becomes larger, reducing the number of matrix elements will help to achieve fewer measurements and more efficient expectation value calculation for matrix entries. 

Inspired by the Quantum Subspace Expansion\cite{mcclean2017hybrid} (QSE) method, our approach neglects the de-excitation operators and inherits the size-consistency from the QEOM method\cite{asthana2023quantum}, the super excitation operator $O^\dagger_n$ becomes
\begin{eqnarray}
\label{eq-bigo2}
O_n^{\dagger} = \sum_{i=0}^{n}Z_i(E^{\alpha}_i)^{\dagger}, 
\end{eqnarray}
where $n$ is the number of basis functions, $Z$ is the expansion coefficient, $\alpha$ indicates the type of the second-quantized excitation operators and the corresponding super de-excitation operator $O_n$ is the complex conjugate of $O_n^\dagger$. In this paper, we restrict $\alpha=1$ to only consider single excitation operators, so
\begin{eqnarray}
\label{eq-num-free-orbital}
(E^{\alpha=1}_i)^{\dagger} = \sum_{j=0}^{m}Z_{ij}a_i^{\dagger}a_j,
\end{eqnarray}
where $m$ labels which atomic orbitals to transition from. For instance, a unit cell of Si has two Si atoms, considering only the atomic $s$ orbital on both Si atoms for the ground state would give $m=2$. This procedure is easily extended to more excitations by including more orbitals.

By substituting Eq.~(\ref{eq-bigo2}) into Eq.~(\ref{eq-qeom2}) and then solving the partial differential equation, we will eventually obtain a new generalized eigenvalue equation in the form of
\begin{eqnarray}
\label{eq-rqeom}
\begin{pmatrix}\mbox{$A$}\end{pmatrix}
\begin{pmatrix}\mbox{$Z$}\end{pmatrix}
= E_{0n}
\begin{pmatrix}\mbox{$B$}\end{pmatrix}
\begin{pmatrix}\mbox{$Z$}\end{pmatrix},
\end{eqnarray}
where
\begin{eqnarray}
\begin{aligned}
\label{eq-rqeom-ab}
\langle A_{ij} \rangle
&= 
\langle\psi_0|[(E^{\alpha}_i)^{\dagger},H,E^{\alpha}_j]|{\psi_0}\rangle, \\
\langle B_{ij} \rangle
&= 
\langle\psi_0|[(E^{\alpha}_i)^{\dagger},E^{\alpha}_j]|{\psi_0}\rangle.    
\end{aligned}
\end{eqnarray}

Given a quantum system with $n$ superoperators and only double excitation operators ($\alpha=2$) are included, the number of fermionic operators in Eq.~(\ref{eq-rqeom}) is $8n^2$, which halves $16n^2$ observables in Eq.~(\ref{eq-qeom3}), leading to speed up the execution when a large number of basis functions is considered. Such algorithms can be efficiently implemented to achieve potential quantum advantages through parallel measurements for the elements of matrices in the problem\cite{ollitrault2020quantum}. In this paper, We shall use the acronym RQEOM to designate the algorithm with fewer number of observables based on Eq.~(\ref{eq-rqeom}).

\begin{figure*}
\includegraphics[scale=0.14]{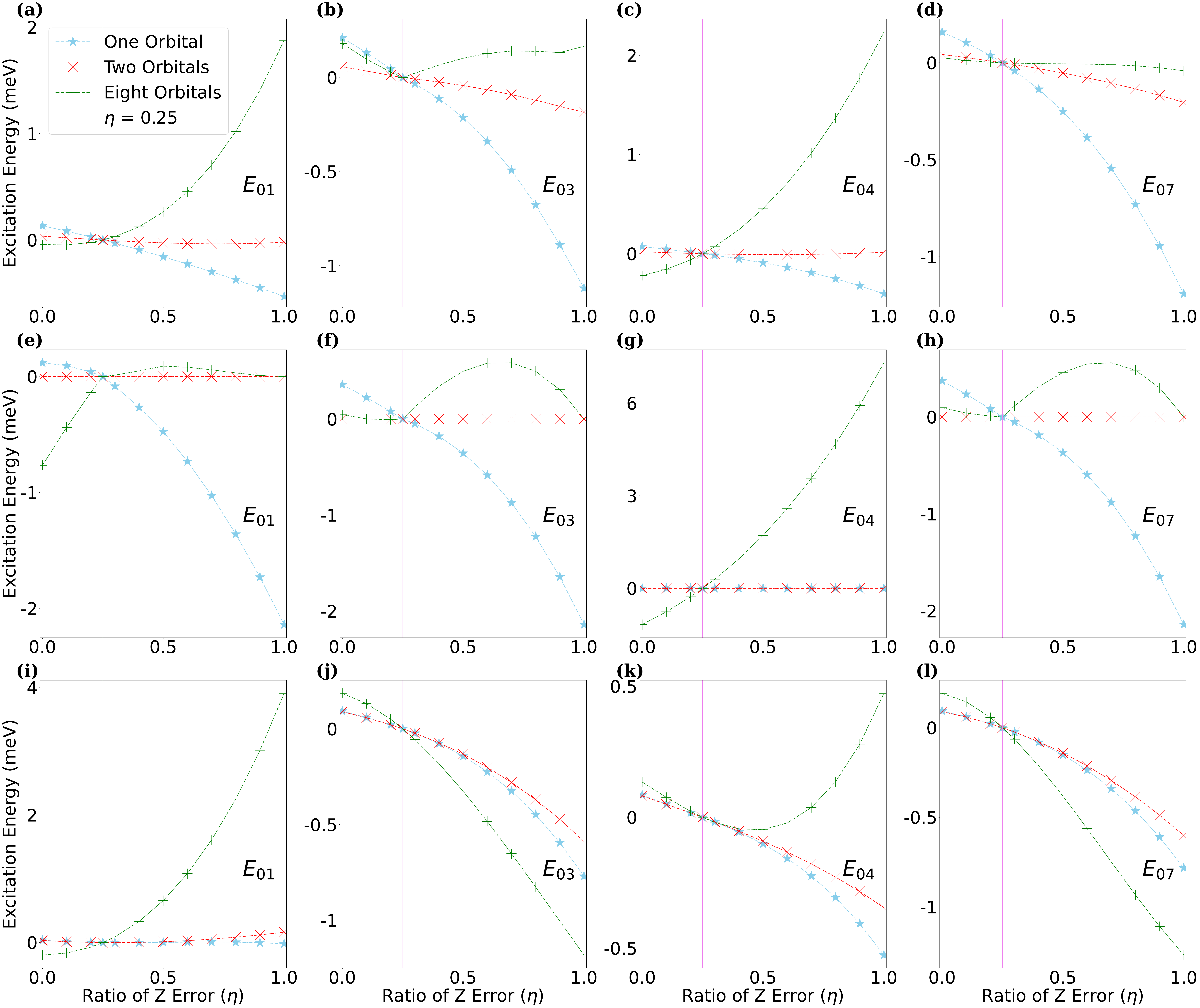}
\caption[Noisy simulation of RQEOM method]{
\label{fig2-1} 
The first ($E_{01}$), third ($E_{03}$), fourth ($E_{04}$) and seventh ($E_{07}$) excitation energy of Si under the biased depolarizing error model ($p = 0.05$) defined in Appendix~\ref{sec-biased-depol-model}.
(a)-(d): The excitation energies of Si at $L$ point.
(e)-(h): The excitation energies of Si at $\Gamma$ point.
(i)-(l): The excitation energies of Si at $U$ point.
The colored crosses depict the excitation energy obtained by RQEOM method with one selected orbital (blue), two selected orbitals (red) and all eight orbitals (green). The exact energy is shown as $0$ on the y-axis, and the vertical pink line illustrates the uniform depolarizing error model ($\eta = 0.25$) where our method is fully immune to it, thus all lines across at point $(0.25, 0)$. 
}
\end{figure*} 

\begin{figure*}
\includegraphics[scale=0.204]{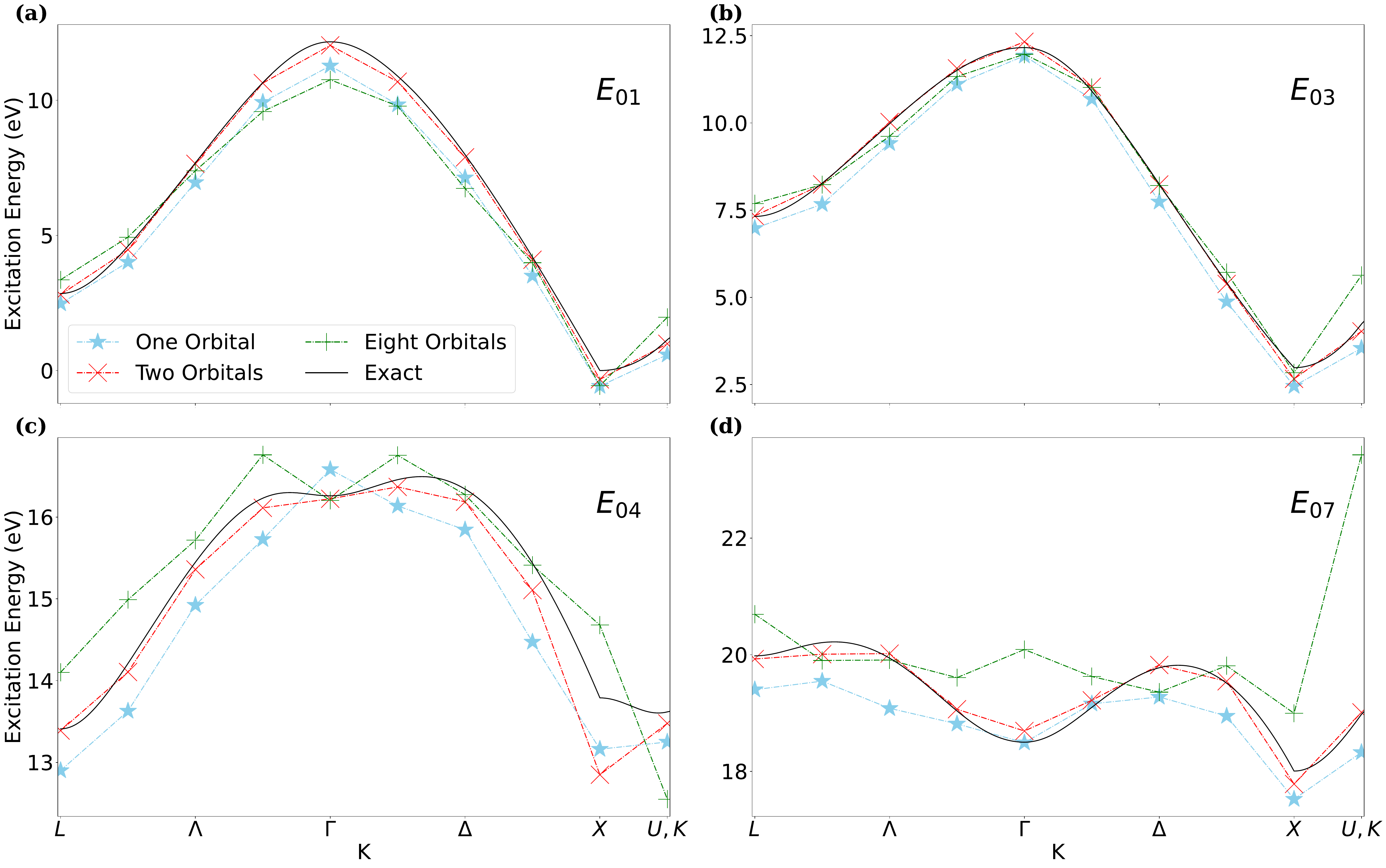}
\caption[Noisy simulation of RQEOM method]{
\label{fig2-2} 
The first ($E_{01}$), third ($E_{03}$), fourth ($E_{04}$) and seventh ($E_{07}$) excitation energy of Si under the hardware noise model derived from $ibm\_hanoi$. Each cross shows the average over 100 trials. The black line represents the exact excitation energy obtained by the classical diagonalization of the TB Hamiltonian.
}
\end{figure*} 

The RQEOM method is summarized as follows. Firstly, the quantum variational algorithm, i.e. VQE, can be used to approximate the ground state $|\psi_0\rangle$ as $|\psi_{\theta_{opt}}\rangle$ and its energy $E_0$ as $E_{min}$. Secondly, a qubit-fermion transformation should be used to map the observables $[(E^{\alpha}_i)^{\dagger},H,E^{\alpha}_j]$ and $[(E^{\alpha}_i)^{\dagger},E^{\alpha}_j]$ from fermions to qubits, in this paper, we choose the classical spectra decomposition to decompose the Hamiltonian into a linear combination of Pauli strings\cite{cerasoli2020quantum}. Lastly, measure the expectation value of the observables for matrix elements in the generalized eigenvalue equation on the state $|\psi_0\rangle$ and solve it classically to obtain the energy gaps $E_{0n}$. The $nth$ excited state energy $E_n$ can be solved by $E_n = E_0 +E_{0n}$. 


\begin{figure*}
\includegraphics[scale=0.204]{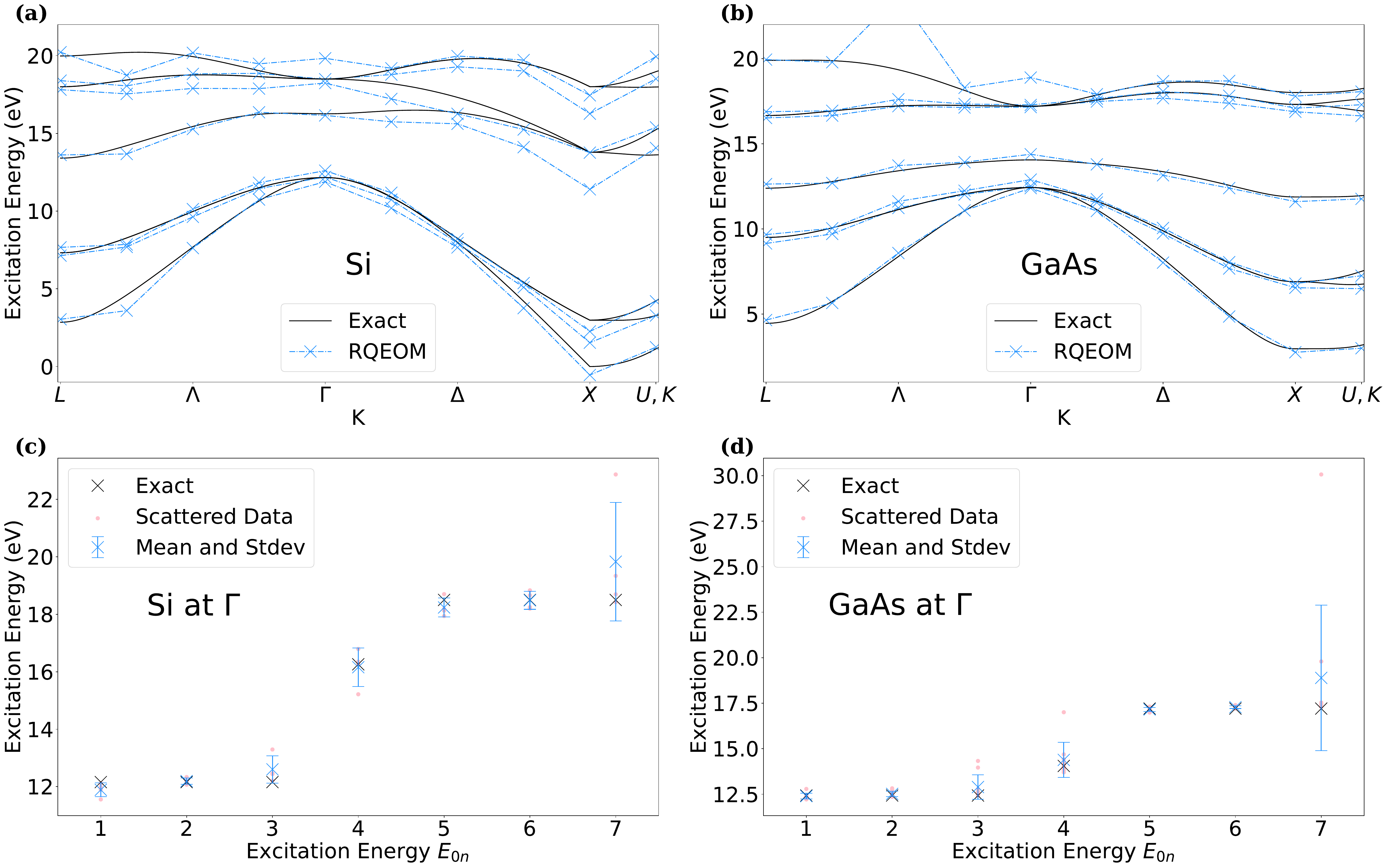}
\caption[Experimental results of RQEOM method on a $ibm_hanoi$]{
\label{fig3} 
(a) The Si excitation energy calculation using RQEOM on $ibm\_hanoi$. Blue crosses are the average of four trials at each k point.
(b) The GaAs excitation energy calculation using RQEOM on $ibm\_hanoi$. Blue crosses are the average of ten trials at each k point.
(c) The scattered experimental data points of Si and the corresponding mean and standard deviation of seven excitation energies at $\Gamma$ point.
(d) The scattered experimental data points of GaAs and the corresponding mean and standard deviation of seven excitation energies at $\Gamma$ point.
}
\end{figure*}

\section{\label{sec-Experiment}Experiments}

In the following sections, we analyze the performance of RQEOM method using Si given two noise models on classical noisy simulators in Section~\ref{subsec-Experiment-Simulation} and calculate the excitation energy of Si and GaAs on a quantum device $ibm\_hanoi$ respectively in Section~\ref{subsec-Experiment-Hardware}. The Hamiltonian of Si and GaAs are approximated by the Tight-Binding (TB) approach\cite{kittel2018introduction, yu2005fundamentals}, where the tetrahedral unit cells form the TB Hamiltonians with the shape of $8\times8$ and can then be further mapped onto a three-qubit Hamiltonian. We use the empirical parameters introduced by Chadi and Cohen\cite{chadi1974tight} for both materials. The open-source package $qiskit$\cite{qiskit} is used for the quantum computing part of our codes. The ground state was solved analytically and a low-depth circuit was designed for direct initialization as the problem size was sufficiently small since the paper focuses on investigating achieving highly accurate excitation energy gaps using RQEOM method given different basis functions for excitation operators, the effect of the ground state should be reduced to not being the major source of the noise.

\subsection{\label{subsec-Experiment-Simulation}Noise Analysis of RQEOM Method}

In this section, we validate the performance of RQEOM method by calculating the excitation energy of Si with three choices of atomic orbitals, $m=1$, $m=2$ and $m=8$ defined in Eq.~\ref{eq-num-free-orbital} under two noise scenarios, where larger m gives more complete descriptions of the system but also brings more observables to measure. Recent studies have shown that errors are biased toward dephasing on superconducting and ion trap qubits\cite{aliferis2009fault, nigg2014quantum}. Thus, for the first scenario, we implement the depolarizing noise model with biased Pauli Z error to test our method under incoherent errors on a statevector simulator, which is a deterministic simulator that contains no sampling noise, the details of the noise model are given in Appendix~\ref{eq-biased-depol}. For the second scenario, we import the modified real hardware noise from $ibm\_hanoi$ to a QASM simulator, which is a non-deterministic model containing shot noise. The selected experimental results are represented in Fig.~\ref{fig2-1} and Fig.~\ref{fig2-2}, and other experimental results are illustrated in Appendix~\ref{sec-exp-detail}.

In Fig.~\ref{fig2-1}, we illustrate the excitation energy calculation of Si with one atomic orbital (blue), two atomic orbitals (red) and eight atomic orbitals (green) using RQEOM approach, we show the first, third, fourth and seventh excitation energies which correspond to the first excited state, the highest valence band, the lowest conduction band and the highest excited state from the leftmost column to the rightmost column respectively, and the Si Hamiltonian is calculated at $L$, $\Gamma$ and $U$ points from the highest row to the lowest row respectively. When $\eta=0.25$, all three dotted colored lines overlap the exact energy, which means the RQEOM method is completely robust to the uniform depolarizing error model. Since the same level of error on every qubit can be regarded as a scalar $p'$ that scales the matrix elements in Eq.~\ref{eq-rqeom} with the same amount of value, $(p'A)(Z)=E_{0n}(p'B)(Z)$, thus, the results of the generalized eigenvalue equation remain the same. The excitation energy shows small changes in the unit of milli-electron volt (meV) when the magnitude of error is set up to $p=0.05$. 


Since the RQEOM method with two orbitals always gives the most robust results under the biased depolarizing error model, we further test our method under a model based on calibration data from a real hardware $ibm\_hanoi$, however, this model cannot fully capture the noise on a real quantum computer, but it is sufficient to be a more complicated noise model before testing on a real NISQ device. In Fig.~\ref{fig2-2} we illustrate the first ($E_{01}$), third ($E_{03}$), fourth ($E_{04}$) and seventh ($E_{07}$) excitation energy of Si calculated by the RQEOM method, since this model is non-deterministic, the sampling noise is also introduced, hence, each cross represents the average of 100 trials. The RQEOM method with $m=2$ (red) always gives the most accurate results compared to the other two lines, whereas the blue and green lines fluctuate around the exact energy and are less close to it than the red line. Since each atom in the unit cell of Si should provide an equal number of atomic orbitals, which gives the most complete and compact description of the system when $m=2$, and the symmetrization could cancel out the effect of noise intrinsically from the perspective of the average of trials. $m=1$ gives an insufficient description of the system, but it provides less number of observables, thus the blue line performs better than the green line at some k points. Increasing the number of orbitals not only increases the number of observables but also introduces extra symmetries, for a non-deterministic model, more observables are easier to be affected by the noise, which leads to the accumulation of noise when there are more than two orbitals, in this paper, we demonstrate the biggest accumulation of noise using $m=8$. Thus, it is vital to find the trade-off between compactness and redundancy of the system when using the RQEOM method.  

\subsection{\label{subsec-Experiment-Hardware}RQEOM Method on IBM Quantum Processor}

We calculate the excitation energy of Si and GaAs using RQEOM method on a NISQ device $ibm\_hanoi$ and show the full experimental results (blue) in Fig.~\ref{fig3}(a) and Fig.~\ref{fig3}(b), where each cross represents the average of four trials and ten trials respectively. No post-selection or noise mitigation method is used in our experiment. The noise varies over time on real devices, thus, the same level of bias for each k point is not expected, but the noise drift can be found in data from $\Gamma$ to $U,K$ in Fig.~\ref{fig3}(a). The lower excitation energies are less affected by noise compared to the higher excitation energies, and most of the degenerate state energies are in between the calculated energies. To investigate such behavior, we scatter the experimental data (pink) of Si and GaAs and the corresponding mean (blue cross) and standard deviation (blue bar) at $\Gamma$ point in Fig.~\ref{fig3}(c) and Fig.~\ref{fig3}(d) respectively, the black crosses shows the exact excitation energies. The first three and last three states of both materials are the degenerate states, when we solve the generalized eigenvalue equation, we obtain each excitation energy by sorting the eigenvalues from the smallest to the largest, since the noise breaks the degeneracy, taking Si as an example, the first three degenerate states at $\Gamma$ have the exact energies of 12.16 eV, the first calculated degenerate state energy (11.89 eV) is always less than the exact energy and the last one (12.60 eV) is always greater than the exact energy, but the middle one (12.19 eV) is almost right on the exact energy, the average of the calculated energies is around 12.67 eV, which is close the exact energy (12.60 eV). For a new material with unknown exact excitation energy, the degenerate states can be found by combining guessing which state should be degenerate from its symmetry with checking the neighbor data points using the averaging process. Besides, since the real hardware also has other types of errors that are hard to identify and simulate in the modified hardware noise model, such as non-Markovian or cross-talk error which is generally the unwanted coupling between qubits\cite{mazda2014telecommunications} instead of the decoherence between qubit and the environment. We observe the data point which is far beyond the standard deviation bar at the seventh excitation energy $E_{07}$ in both Fig.~\ref{fig3}(c) and Fig.~\ref{fig3}(d), which leads to a large mean value of 19.83 eV and 18.89 eV. However, such sparsely distributed data points can be regarded as statistical outliers when more trials are obtained. For instance, the data cluster can be observed in the seventh excitation energy $E_{07}$ in Fig.~\ref{fig3}(d) due to the increasing number of trials for GaAs, the mean value would be close to the exact energy without the outliers, thus, post data selection methods can be used to filter out those outliers based on the confidence of each data point to give a more precise result. For the fourth excitation energy $E_{04}$ in Fig.~\ref{fig3}(c), averaging the experimental data is sufficient enough to obtain an accurate energy (16.16 eV) due to the hardware noise appears to result in random noise around the exact energy (16.26 eV), but more trials are needed for larger the quantum system to give better robustness. Error mitigation techniques can be applied to raw experimental data for future research. The chip geometry of $ibm\_hanoi$ and the characteristics of each qubit are shown in Appendix.~\ref{sec-device-charac}.


\section{\label{sec-Conclusions}Conclusions}

In this work, we reduce the number of observables and the required number of observables in QEOM method and propose the RQEOM method to directly calculate the excitation energy of a quantum system. We investigate the performance of our method by calculating the excitation energy of Si on a biased depolarizing error model, a hardware noise model, then we implement our method on a NISQ device to obtain the excitation energies of Si and GaAs. The RQEOM method is fully robust to the uniform depolarizing error, finding the suitable orbital complexity in Eq.~\ref{eq-num-free-orbital} could increase the fidelity of energies. There is a natural error mitigation in the algorithm itself and a large number of trials results in the average approaching the exact energy. We also point out that future directions could focus on applying statistical post-selection methods and error mitigation algorithms to test the performance of RQEOM method.


\begin{acknowledgments}

We acknowledge the use of IBM Quantum services for this work. The views expressed are those of the authors, and do not reflect the official policy or position of IBM or the IBM Quantum team.

\end{acknowledgments}


%


\appendix

\section{\label{sec-biased-depol-model}Biased Depolarizing Error Model}

The biased depolarizing quantum channel on a single qubit is defined as
\begin{eqnarray}
\begin{aligned}
\label{eq-biased-depol}
\mathcal{E}(\rho) = (1 - p) \rho + p\sum_ \sigma r_\sigma \sigma^\dagger \rho \sigma,
\end{aligned}
\end{eqnarray}
where $p$ is the probability of a depolarizing error occurring on a single qubit, $\sigma$ is the single Pauli error channel, $\sigma \in \{I, X, Y, Z\}$, $r_\sigma$ defined as the proportion of a $\sigma$ error to all the noise and parametrized by the parameter $\eta$. In the case of a $Z$-biased error model, $r_I = r_X = r_Y = (1-\eta)/3$ and $r_Z = \eta$, where $\eta$ is the ratio of Pauli $Z$ error parametrizes the complete model, $0 \leq \eta \leq1$, when $\eta=1/4$, this gives a general depolarizing error model; when $\eta=1$, the model corresponds to a pure $Z$ error model.

\section{\label{sec-exp-detail}Supplementary Experimental Results}

\begin{figure*}
\includegraphics[scale=0.14]{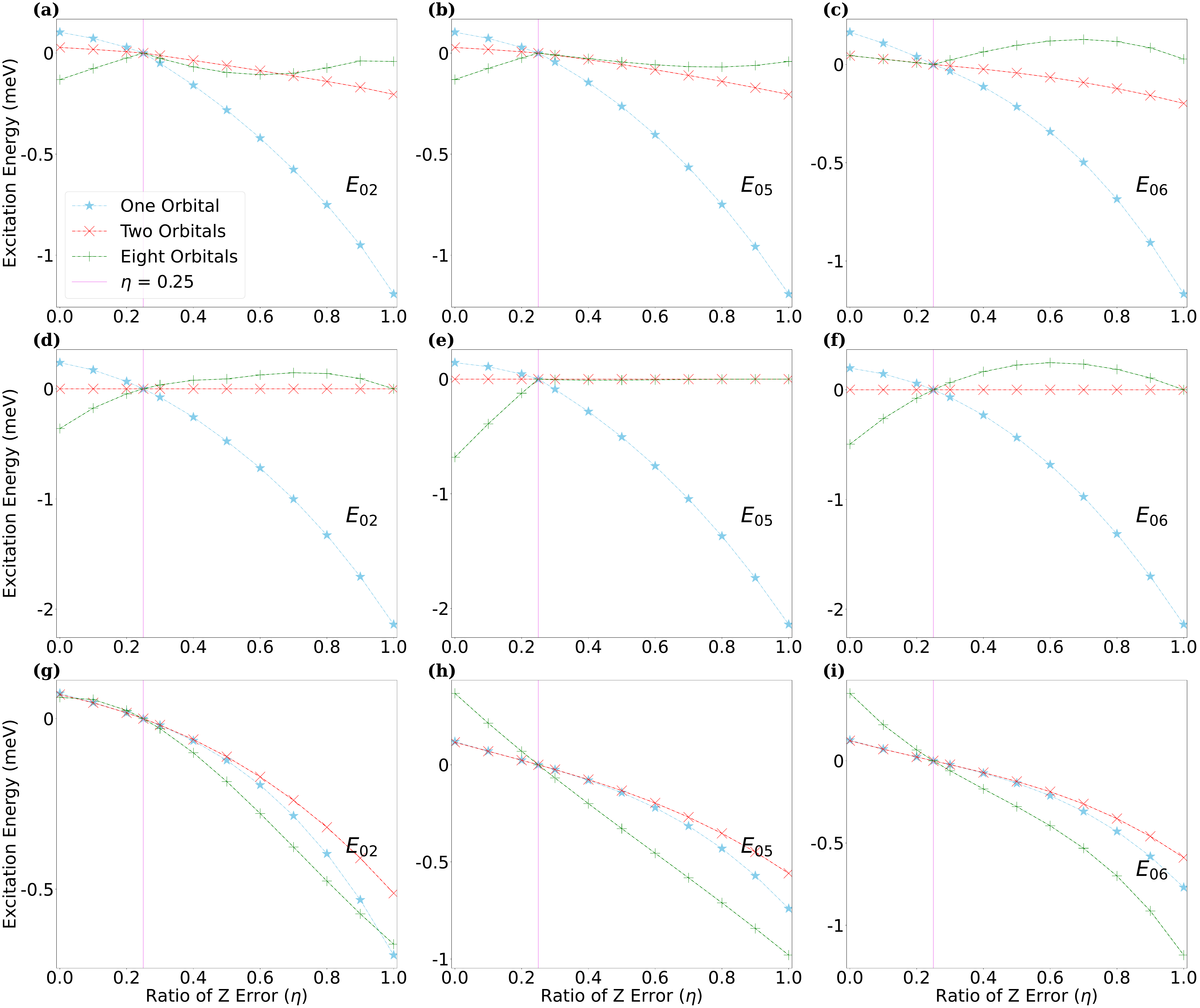}
\caption[Noisy simulation of RQEOM method]{
\label{fig-appendixB1} 
The second ($E_{02}$), fifth ($E_{05}$) and sixth ($E_{06}$) excitation energy of Si under the biased depolarizing error model ($p = 0.05$) defined in Appendix~\ref{sec-biased-depol-model}.
(a)-(c): The excitation energies of Si at $L$ point.
(d)-(f): The excitation energies of Si at $\Gamma$ point.
(g)-(i): The excitation energies of Si at $U$ point.
The colored crosses depict the excitation energy obtained by RQEOM method with one selected orbital (blue), two selected orbitals (red) and all eight orbitals (green). The exact energy is shown as $0$ on the y-axis, the vertical pink line illustrates the uniform depolarizing error model ($\eta = 0.25$) where our method is fully immune to it, thus all lines across at point $(0.25, 0)$. 
}
\end{figure*}

The supplementary experimental results of the biased depolarizing error simulation of the RQEOM method are illustrated in Fig.~\ref{fig-appendixB1}, which shows the second, fifth and sixth excitation energy on each column respectively under three orbital complexities. Similar to the results in Fig.~\ref{fig2-1}, Choosing two atomic orbitals gives the best performance compared to the other two scenarios, since the symmetry of the system obtained by the suitable orbitals naturally mitigates the effect of the biased depolarizing noise.

\section{\label{sec-device-charac}Quantum Device Characteristics}

\begin{figure*}
\includegraphics[scale=0.3]{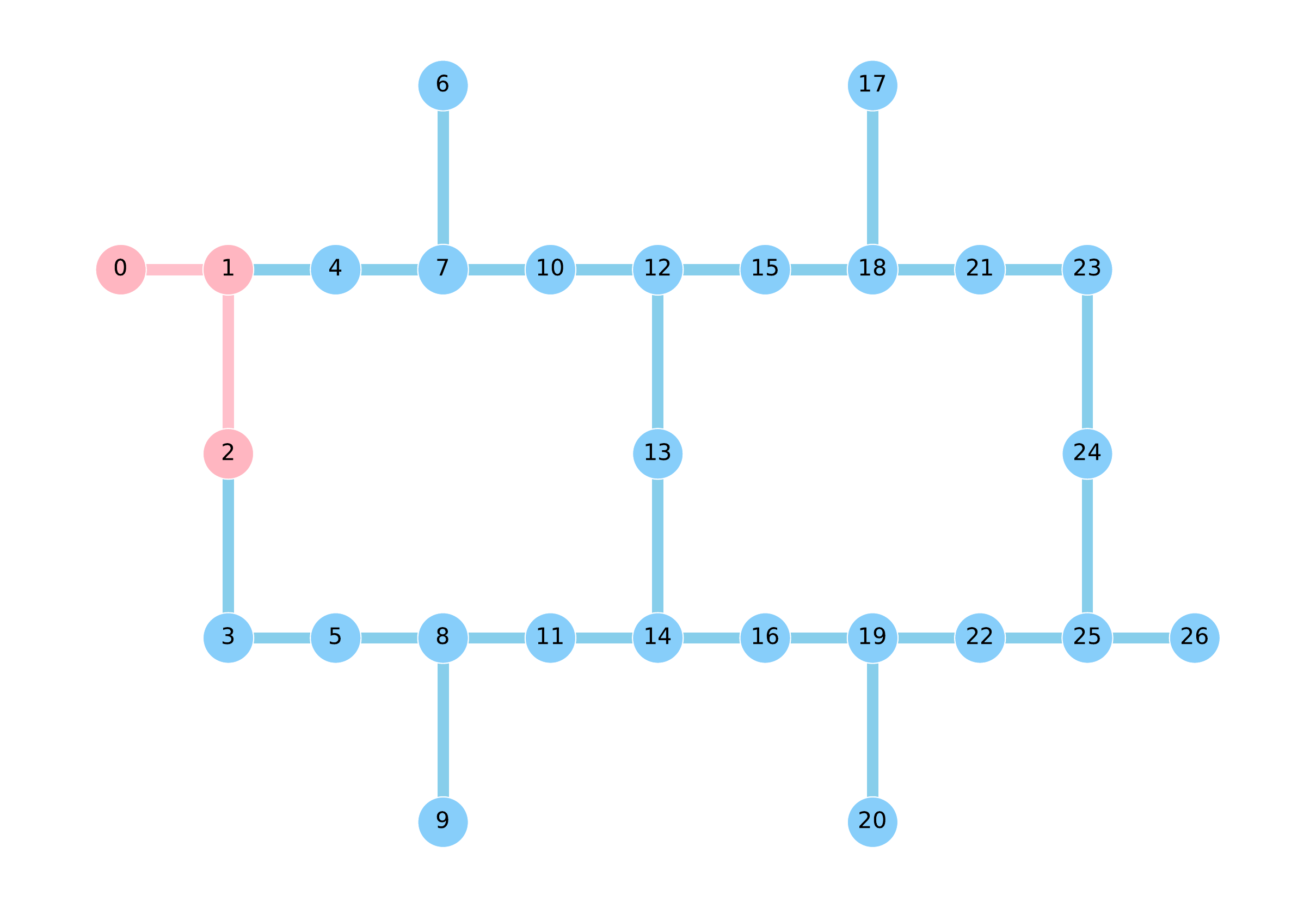}
\caption[Chip geometry of $ibm_hanoi$]{
\label{fig-appendixB2} 
Chip geometry of $ibm\_hanoi$, colored circles and edges denote qubits and their connections respectively, pink nodes and edges are used in Section~\ref{subsec-Experiment-Hardware}.
}
\end{figure*}

We illustrate the chip geometry of $ibm\_hanoi$ in Fig.~\ref{fig-appendixB2}, where the pink nodes and edges denote the qubits and connections that are used in our experiments. The corresponding noise spectroscopy of the qubits is shown in Table~\ref{table-device-charac}, where the data was collected during the execution of our experiments in Section~\ref{subsec-Experiment-Hardware}. The calibration runs every day on each IBM quantum device and the status of qubits varies over time, which makes the data different from the other days. Besides, other types of noise cannot be precisely characterized on real devices, thus, using the collected noise cannot guarantee obtaining the same experimental results as on real devices.

\afterpage{
\clearpage
\thispagestyle{empty}
\centering
\begin{sidewaystable}
\caption{\label{table-device-charac} 
Device Characteristics on $ibm\_hanoi$}
\adjustbox{scale=0.86,right=25.16cm,vspace*=0pt -9cm}{
\begin{tabular}{|c|c|c|c|c|c|c|c|c|c|c|c|c|c|c|c|c|c|c|c|c|c|c|c|c|c|c|c|}
\hline
Qubit                 & 0      & 1      & 2      & 3     & 4     & 5      & 6      & 7      & 8      & 9      & 10     & 11     & 12     & 13     & 14     & 15     & 16     & 17     & 18     & 19     & 20     & 21     & 22     & 23     & 24     & 25     & 26    \\ \hline
$T_1$ ($\mu s$)               & 177.12 & 130.62 & 197.54 & 96.79 & 77.26 & 153.66 & 86.48  & 145.77 & 140.89 & 105.12 & 120.97 & 150.01 & 97.34  & 151.12 & 198.13 & 170.77 & 154.38 & 126.65 & 126.64 & 148.77 & 121.31 & 140.81 & 169.05 & 146.44 & 112.27 & 187.65 & 59.95 \\ \hline
$T_2$ ($\mu s$)               & 324.59 & 127.81 & 248.39 & 33.52 & 14.53 & 278.55 & 102.42 & 174.46 & 275.91 & 141.74 & 260.35 & 195.94 & 121.88 & 228.55 & 26.77  & 36.80  & 294.84 & 83.22  & 144.13 & 162.47 & 93.73  & 40.90  & 128.43 & 84.51  & 37.09  & 91.16  & 31.56 \\ \hline
Frequency (GHz)       & 5.04   & 5.16   & 5.26   & 5.10  & 5.07  & 5.21   & 5.02   & 4.92   & 5.03   & 4.87   & 4.82   & 5.16   & 4.72   & 4.96   & 5.05   & 4.92   & 4.88   & 5.22   & 4.97   & 5.00   & 5.10   & 4.84   & 4.92   & 4.92   & 4.99   & 4.81   & 5.02  \\ \hline
Readout error (\%)    & 0.75   & 0.97   & 1.62   & 0.66  & 0.96  & 0.93   & 1.83   & 1.10   & 0.95   & 0.65   & 1.33   & 7.79   & 1.63   & 5.04   & 0.82   & 2.51   & 1.01   & 1.86   & 1.42   & 0.86   & 0.92   & 0.79   & 1.23   & 1.19   & 0.76   & 1.05   & 0.93  \\ \hline
Prob meas0 prep1 (\%) & 0.90   & 1.06   & 1.98   & 0.60  & 1.24  & 1.50   & 2.36   & 1.30   & 1.36   & 1.04   & 1.64   & 8.00   & 1.74   & 4.80   & 1.04   & 2.20   & 1.04   & 2.04   & 1.98   & 0.96   & 1.02   & 1.04   & 1.50   & 1.30   & 0.82   & 1.48   & 0.94  \\ \hline
Prob meas1 prep0 (\%) & 0.60   & 0.88   & 1.26   & 0.72  & 0.68  & 0.36   & 1.30   & 0.90   & 0.54   & 0.26   & 1.02   & 7.58   & 1.52   & 5.28   & 0.60   & 2.82   & 0.98   & 1.68   & 0.86   & 0.76   & 0.82   & 0.54   & 0.96   & 1.08   & 0.70   & 0.62   & 0.92  \\ \hline
$CX_{q_0,q_n}$ error (\%)    &        & 0.61   &        &       &       &        &        &        &        &        &        &        &        &        &        &        &        &        &        &        &        &        &        &        &        &        &       \\ \hline
$CX_{q_1,q_n}$ error (\%)    & 0.61   &        & 0.30   &       & 0.35  &        &        &        &        &        &        &        &        &        &        &        &        &        &        &        &        &        &        &        &        &        &       \\ \hline
$CX_{q_2,q_n}$ error (\%)    &        & 0.30   &        & 0.85  &       &        &        &        &        &        &        &        &        &        &        &        &        &        &        &        &        &        &        &        &        &        &       \\ \hline
$CX_{q_3,q_n}$ error (\%)    &        &        & 0.85   &       &       & 0.42   &        &        &        &        &        &        &        &        &        &        &        &        &        &        &        &        &        &        &        &        &       \\ \hline
$CX_{q_4,q_n}$ error (\%)    &        & 0.35   &        &       &       &        &        & 1.27   &        &        &        &        &        &        &        &        &        &        &        &        &        &        &        &        &        &        &       \\ \hline
$CX_{q_5,q_n}$ error (\%)    &        &        &        & 0.42  &       &        &        &        & 100.00 &        &        &        &        &        &        &        &        &        &        &        &        &        &        &        &        &        &       \\ \hline
$CX_{q_6,q_n}$ error (\%)    &        &        &        &       &       &        &        & 1.10   &        &        &        &        &        &        &        &        &        &        &        &        &        &        &        &        &        &        &       \\ \hline
$CX_{q_7,q_n}$ error (\%)    &        &        &        &       & 1.27  &        & 1.10   &        &        &        & 1.09   &        &        &        &        &        &        &        &        &        &        &        &        &        &        &        &       \\ \hline
$CX_{q_8,q_n}$ error (\%)    &        &        &        &       &       & 100.00 &        &        &        & 0.66   &        & 0.45   &        &        &        &        &        &        &        &        &        &        &        &        &        &        &       \\ \hline
$CX_{q_9,q_n}$ error (\%)    &        &        &        &       &       &        &        &        & 0.66   &        &        &        &        &        &        &        &        &        &        &        &        &        &        &        &        &        &       \\ \hline
$CX_{q_10,q_n}$ error (\%)   &        &        &        &       &       &        &        & 1.09   &        &        &        &        & 0.59   &        &        &        &        &        &        &        &        &        &        &        &        &        &       \\ \hline
$CX_{q_11,q_n}$ error (\%)   &        &        &        &       &       &        &        &        & 0.45   &        &        &        &        &        & 0.85   &        &        &        &        &        &        &        &        &        &        &        &       \\ \hline
$CX_{q_12,q_n}$ error (\%)   &        &        &        &       &       &        &        &        &        &        & 0.59   &        &        & 0.62   &        & 0.92   &        &        &        &        &        &        &        &        &        &        &       \\ \hline
$CX_{q_13,q_n}$ error (\%)   &        &        &        &       &       &        &        &        &        &        &        &        & 0.62   &        & 0.51   &        &        &        &        &        &        &        &        &        &        &        &       \\ \hline
$CX_{q_14,q_n}$ error (\%)   &        &        &        &       &       &        &        &        &        &        &        & 0.85   &        & 0.51   &        &        & 1.28   &        &        &        &        &        &        &        &        &        &       \\ \hline
$CX_{q_15,q_n}$ error (\%)   &        &        &        &       &       &        &        &        &        &        &        &        & 0.92   &        &        &        &        &        & 1.34   &        &        &        &        &        &        &        &       \\ \hline
$CX_{q_16,q_n}$ error (\%)   &        &        &        &       &       &        &        &        &        &        &        &        &        &        & 1.28   &        &        &        &        & 0.37   &        &        &        &        &        &        &       \\ \hline
$CX_{q_17,q_n}$ error (\%)   &        &        &        &       &       &        &        &        &        &        &        &        &        &        &        &        &        &        & 0.55   &        &        &        &        &        &        &        &       \\ \hline
$CX_{q_18,q_n}$ error (\%)   &        &        &        &       &       &        &        &        &        &        &        &        &        &        &        & 1.34   &        & 0.55   &        &        &        & 0.37   &        &        &        &        &       \\ \hline
$CX_{q_19,q_n}$ error (\%)   &        &        &        &       &       &        &        &        &        &        &        &        &        &        &        &        & 0.37   &        &        &        & 100.00 &        & 0.64   &        &        &        &       \\ \hline
$CX_{q_20,q_n}$ error (\%)   &        &        &        &       &       &        &        &        &        &        &        &        &        &        &        &        &        &        &        & 100.00 &        &        &        &        &        &        &       \\ \hline
$CX_{q_21,q_n}$ error (\%)   &        &        &        &       &       &        &        &        &        &        &        &        &        &        &        &        &        &        & 0.37   &        &        &        &        & 0.82   &        &        &       \\ \hline
$CX_{q_22,q_n}$ error (\%)   &        &        &        &       &       &        &        &        &        &        &        &        &        &        &        &        &        &        &        & 0.64   &        &        &        &        &        & 0.74   &       \\ \hline
$CX_{q_23,q_n}$ error (\%)   &        &        &        &       &       &        &        &        &        &        &        &        &        &        &        &        &        &        &        &        &        & 0.82   &        &        & 2.51   &        &       \\ \hline
$CX_{q_24,q_n}$ error (\%)   &        &        &        &       &       &        &        &        &        &        &        &        &        &        &        &        &        &        &        &        &        &        &        & 2.51   &        & 1.64   &       \\ \hline
$CX_{q_25,q_n}$ error (\%)   &        &        &        &       &       &        &        &        &        &        &        &        &        &        &        &        &        &        &        &        &        &        & 0.74   &        & 1.64   &        & 0.75  \\ \hline
$CX_{q_26,q_n}$ error (\%)   &        &        &        &       &       &        &        &        &        &        &        &        &        &        &        &        &        &        &        &        &        &        &        &        &        & 0.75   &       \\ \hline
\end{tabular}
}
\end{sidewaystable}
}

\end{document}